# Intercontinental quantum liaisons between entangled electrons in ion traps of thermoluminescent crystals

Robert Desbrandes (Louisiana State University) and Daniel L. Van Gent (Oklahoma State University)

**Abstract**

The experiments reported in this paper were carried out with space-separated entangled TLD crystals in Baton Rouge, Louisiana (USA) and Givarlais (France) distant of 8,182 km. Samples of doped lithium fluoride, an excellent TLD material used industrially have been irradiated together at one location in order to create entangled traps in spatially collocated TLD chips via simultaneous Bremsstrahlung irradiation of a medical accelerator on spatially collocated pairs of TLD chips. One of the chips was then mailed to Baton Rouge and its entangled counterpart was kept in Givarlais. Thermally heating the sample located in Baton Rouge has produced in the corresponding entangled sample measured under a photomultiplier in Givarlais at ambient temperature, correlated signals while the TLD temperature was increased and then allowed to decrease by turning off the TLD heating oven in Baton Rouge. The instant that maximum temperature of the TLD is attained in Baton Rouge corresponds exactly with the instant of maximum correlation of PMT signal recorded in Givarlais

## 1 - Introduction

Numerous experiments have been reported lately concerning photons entanglement. Entanglement has been predicted by Quantum Mechanics in the late 1920. In 1935 Einstein, Podolski and Rosen wrote a paper [1] in which they question the validity of the entanglement concept resulting from the theory and suggested the existence of hidden variables to explain entanglement. In 1962, J. S. Bell [2], mathematically showed that experiments could be conducted to validate the Quantum Mechanics prediction. Then in 1980 [3], A. Aspect experimentally showed that, applying Bell criteria, entanglement of photons was a phenomenon ruled by Quantum Mechanics. In the1990-2000, several experimentalists showed that entangled photons generated by non-linear crystals could stay entangled for distances up to 10 km [4]. The instant of the entanglement collapse due to the polarization measurement of one of the photons caused the immediate determination of the polarization of the other photon in agreement with Quantum Mechanics. Teleportation experiments in which one photon can be reproduced while being transported by a couple of entangled photons have also been carried out [5]. Entanglement swapping consisting in the transfer of entanglement from a set of particles to another has been studied theoretically [6] and demonstrated experimentally. [7]. The authors have carried out some work with gamma emitted from Cobalt 60 and from Bremsstrahlung gamma emitted by a linear accelerator. These experiments show that such gamma are entangled since they can swap their entanglement to metastable nuclei while exciting them. The nuclei entanglement is then verified by a shorter initial half-life of the excited nuclei during their natural deexcitation [8]. The authors, then, irradiated two samples together, separated them spatially, and verified that deexciting one by X-rays stimulation [9] would cause the deexcitation of the other sample [10] in conformity with Quantum Mechanics. Some work was also carried out with photoluminescent samples [11] in the same domain. The present work is a continuation using traps in thermoluminescent samples for entanglement swapping with the entangled X-rays or gamma emitted by the sources mentioned previously.

## 2 - Brief entanglement theory

Quantum Mechanics teaches us that when two particles are emitted simultaneously or quasi-simultaneously by the same entity, they are entangled. They have then a common wave function that can be written as:

$$|\Psi>_{AB} = 1/\sqrt{2} \,[|0>_A (x) |1>_B - |1>_A (x) |0>_B ]$$

(x) meaning a tensor product.

We see here, that when particle A is measured in state $|0>_A$, then the wave function collapses and particle B immediately takes state $|1>_B$.
This theory has been proven in the experiments mentioned above.
In the case of the present work, the state of particle A is not measured, it is forced by stimulation to go from state $|1>_A$ to state $|0>_A$, consequently, particle B has to go to state $|0>_B$, thus explaining the experiments reported in papers [8, 10] as well as the present experiments.
It is interesting to note that the wave function collapse does not take any account of the spatial position of the particles A and B, which suggest that the phenomenon of collapse, in the case of spatially separated particles, is instantaneous and independent of the distance and of the mediums containing the particles.

## 3 - Brief thermoluminescence theory

Thermoluminescence occurs in crystals that contains imperfections, or impurity atoms, or atoms of doping. Such crystals have the property of storing the effect of the irradiation with X-rays or gamma rays. This storage can last for years and is used for geological and archeological dating with natural or pottery type materials. Artificial crystals are used for thermoluminescent dosimetry (TLD) to determine the exposure to ionizing radiations. When the crystals are heated, the energy is released in the form of light. In the reported experiment, dosimetry type crystals have been used.
The thermoluminescence is explain in the band theory as follows [12]:

- The ionization of network atoms, due to X-ray or Gamma radiations free some electrons from the valence band. Some holes are formed, and the electrons get into the energetic continuum of the conduction band.
- These electrons are then captured by the traps formed by the imperfections, or impurity atoms, or atoms of doping, in the forbidden band of the crystal network. They are then in a metastable state.
- Such a metastable state can, according to the type of crystalline material last from a very short time to thousands of years.
- Heat or optical energy applied to the crystal allows the electrons to leave the traps. They, then, come back in the valence band while emitting some light and producing the thermoluminescence phenomenon.

The traps in the crystals need various calorific or optical energy to liberate the electrons according to their "depth" in terms of energy. "Shallow" traps will empty at low temperature, for example 140°C, "deep" traps will require higher temperature, for example 240°C. For the entangled electrons, as it will be explain below, the phenomenon of emptying of the traps seems much more complex.

## 4 - Bremsstrahlung gamma entanglement

Bremsstrahlung or "braking radiation" is the electromagnetic radiation produced by the acceleration of a charge particle, for example an electron, that is deflected by another charged particle such as an atom nucleus. It was discovered in 1888 by Nikola Tesla. The Bremsstrahlung effect is mostly used today for the radiations that are produced by the deceleration of a charged particle when deflected by another charged particle.

In the experiments presented, the radiation is due to a beam of high energy electron aimed at a target of Tungsten. The efficiency of such a process is given by:
$$\eta = C\,U\,Z$$
C is a coefficient, U is the voltage, and Z the atomic number.
For the compact linear accelerator used in the experiment, U = 6 MeV, and Z = 184
$$\eta = 40\%$$
Due to the large amount of heat produced, such an accelerator cannot work permanently, and the large irradiations must be done in several doses.
In heavy atoms such as Tungsten, a wide spectrum of photons is produced since one electron gives off, as an average, many photons. This spectrum is maximal for about 1.5 MeV when the electrons are accelerated at 6 MeV [13]. The photons are produced in a cone with an apex angle depending on energy. For example: 5° for 35 MeV, and about 30° for 6 MeV.

Since one electron produces several photons instantaneously, such photons are entangled according to Quantum Mechanics. Being statistically spread in a cone, entangled photons can hit several samples, swap their entanglement to nuclei or to atom electrons. Electron accelerators are thus efficient tools for irradiating thermoluminescent materials where entangled electrons can be stored.

## 5 - Entanglement swapping

In the last few years many articles have been published on entanglement swapping, mostly for applications in computers. The conventional type of entanglement swapping is shown in Figure 1.

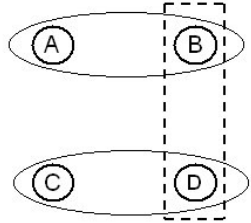

Figure 1 – Entanglement swapping between two entangled pairs of particles. (After reference 7).

The theory shows that if a measurement is made simultaneously on element (B) and (D) of the entangled pairs (A) (B) and (C) (D), the entanglement on pairs (A) (B) and (C) (D) collapses, but the elements (A) and (C) become entangled although they have never been in contact before.

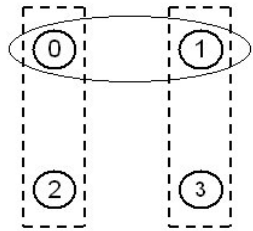

Figure 2 – Entanglement swapping between two entangled gamma and two electrons.

In the case reported here, Figure 2 schematically shows two entangled gamma (0) and (1), interacting simultaneously with two electrons (2) and (3) in a crystal. It will cause the entanglement of the electrons (2) and (3) and the entanglement collapse between (0) and (1). These entangled electrons are then captured in the crystal traps and may stay as such in the traps for months or years at ambient temperature.

## 6 - Storing entanglement

Recent work with electrons in magnetic materials show that at very low temperature the electron motion does not stop [14]. Electron spins are entangled in pairs and their innate bond with one another keep them from freezing, contrary to what classical physics predicts [12].
In a simplified theory of luminescence the probability of release of the trapped electrons is given by Arrhenius equation:
$$p = A \exp(-E/kT)$$
in which,   A is a constant
            p, the probability of release of the trapped electrons,

E, the trap depth (energy),
k, the Boltzmann constant,
T, the absolute temperature.

This equation indicates that in the case of non-entangled electrons in the traps of themoluminescent materials, the traps start emptying as soon as the temperature generates phonons with an energy sufficient to trigger the emptying of the traps. For example, this energy at 140 °C is $kT = 5.37 \times 10^{-21}$ J. ( $k = 1.38 \times 10^{-23}$ J/K). The traps keep emptying as the temperature increases. This emptying may be slow and particular techniques are used to get a complete "annealing", however as kT is much greater than E during annealing, the traps empty reasonably fast. This phenomenon is seen experimentally when measuring the thermally stimulated luminescence glow of CLINAC irradiated TLD chips. When the chips are subjected to a second thermal cycle, the signal response is only about 2 to 3% of the original stimulated luminescence measured during the first thermal cycle. Complete annealing for LiF200 TLDs requires at least 10 minutes at 250°C for complete emptying most of the traps.. For LiF100 TLD's according to Reference 12, annealing requires a sustained temperature of 400°C for at least one hour.

Entangled electrons, as the experiments show, do not appear to exit from the traps as the temperature increases except at very discrete and narrow characteristic trap emptying temperatures. Consequently, shallow trap emptying at full intensity, for example at 140°C, is observed again as the temperature decreases or "turns around" during cooling after the initial temperature excursion increase up to 250°C or more. This phenomenon can be witnessed after several temperature excursion cycles up to 250°C or more as is shown in the experiments.

Furthermore the entangled electrons seems to be resistant to decoherence, that is the collapse of their entanglement, since samples, co-irradiated several months prior to the experiments reported here, still gave intense signals. An extremely large number of single electrons in one crystal are entangled with electrons in the other crystal and "stored" in relative decoherence free space of impurity traps of the crystals. It appears that the ion traps are behaving very similarly to QED quantum cavities.

**7 - Laboratory experiments**

The two types of chips used in the experiment are lithium fluoride LiF100M and LiF 200A. Their characteristics are given in Table 1. A photograph of the chips can be seen in Figure 3.

| Name | Chemistry | Shape | Size (mm) | Usage (Gy) | Glow peaks (°C) |
|---|---|---|---|---|---|
| LiF100M | LiF:Mg,Ti | Square | 3.2-3.2-0.9 | $50 \times 10^{-6}$ to 500 | 137-170-190-210 |
| LiF200A | LiF:Mg,Cu,P | Cylindrical | 4.5 Dia 0.8 | $0.5 \times 10^{-6}$ to 12 | 145 – 232 |

Table 1: Main characteristics of LiF thermoluminescent chips.
(Data from references 12 and 15)

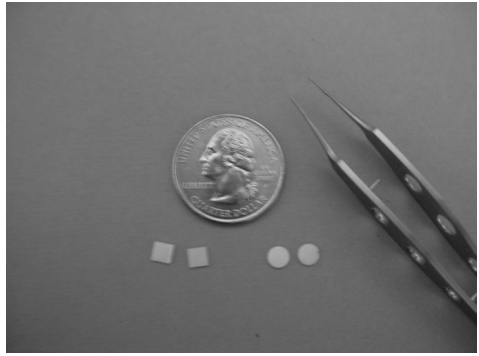

Figure 3: Chips LiF 100 are shown to the left, chips LiF 200 to the right, the coin is a quarter dollar.

The laboratory experimental setup is shown in Figure 4. A commercial oven equipped with a recording thermometer and a thermostat allowing for cutoff at the chosen maximum temperature and no restarting. The master chip is placed between two aluminum foils.

On the left of the figure, and 4 meters away, a photomultiplier is mounted in a thick steel cylinder and connected to a computer for data recording. The slave chip is placed 10 mm away from the 25 mm diameter photosensitive face of the photomultiplier.

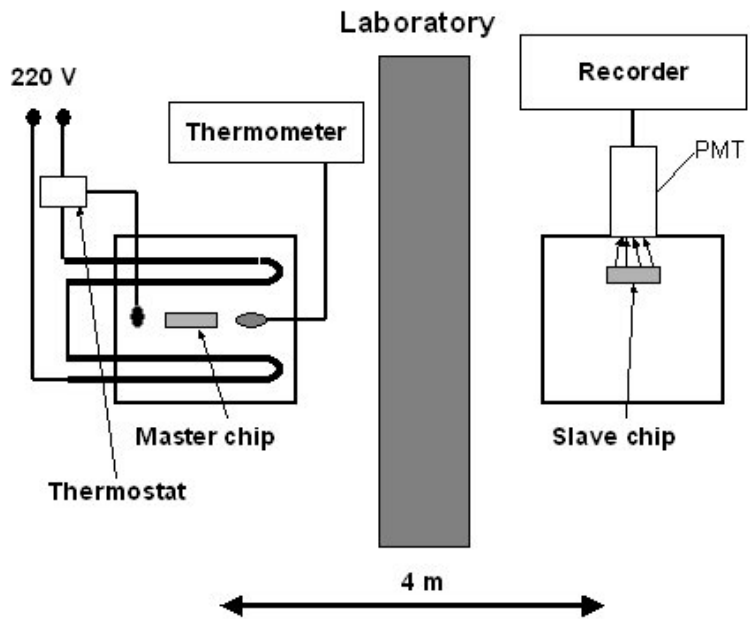

Figure 4 – Laboratory setup

While the oven is rising and then decreasing in temperature, a recording versus time is done as shown in Figure 5. In this typical recording one sees a repeat of the glow curve as the temperature decreases. Since the rate of decrease in temperature is lower, the peaks are larger.

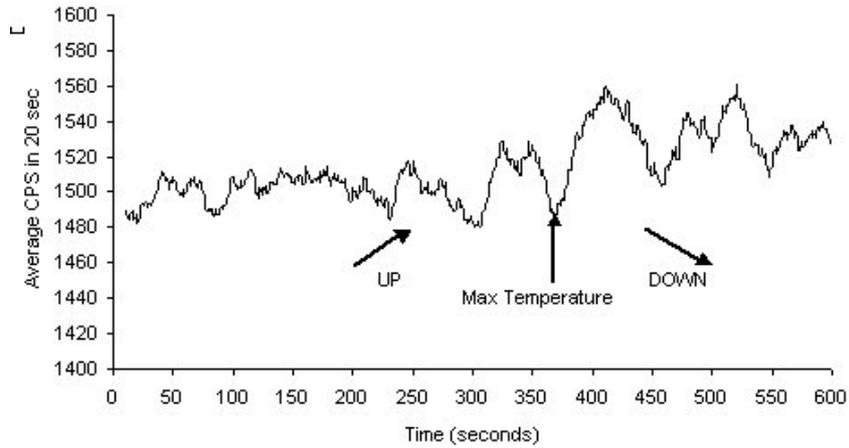

Figure 5 - Test 2, LiF200, 2 Gray, PMT 2, with heat, 2006
Signal on Slave Chip (not heated). Master chip heated 4 m away

Taking into account the different rate of temperature variation, a curve of the glow curves up and down versus temperature can be computed as shown in Figure 6. For clarity, the down temperature curve, after the temperature rate correction, has been shifted slightly upwards. An excellent correlation coefficient of 0.80 between the glow curves up and down in temperature shows the particular behavior of the entangled electrons in the traps.

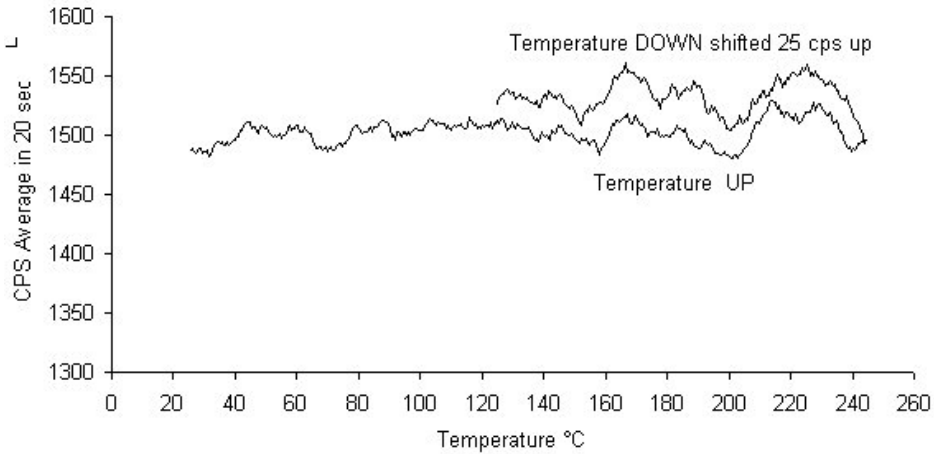

Figure 6 - Test of chip 2, LiF200, 2 Gray, PMT 2, with heat, 2006
Signal on Slave Chip (not heated). Master chip heated 4 m away.

## 8 - Transatlantic experiments

Transatlantic experiments were conducted to test the independency of the entanglement phenomenon with distance. In a set of two chips irradiated together, the master chip was mailed to Baton Rouge, Louisiana, USA, to be heated there in a domestic oven similar to the oven used in the laboratory experiment in Givarlais, France. Givarlais is located at 8182 km from Baton Rouge along a great circle and at 7630 km in straight line. Figure 7 schematically represent the setup of the test. By telephone the start of the heating cycle was transmitted from Baton Rouge to Givarlais. A recording of the glow curve of the slave chip was made in Givarlais and the data computed versus temperature taking into account the different rate of rise and decrease of the temperature in Baton Rouge. The data were then shifted to get the best correlation coefficient between the glow curve when the temperature was going up an down. A small shift was generally required due to the ambient temperature of the garage were the oven containing the master chip was operating. The time of reversal of the temperature, checked for each test within one second of the time measured in Baton Rouge.

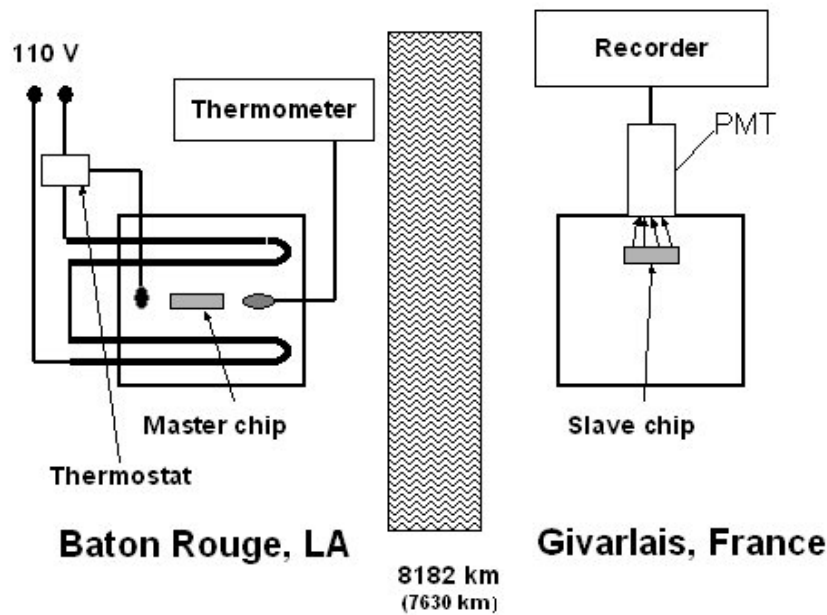

Figure 7 -  Setup for the  Intercontinental test  between Baton Rouge, LA (USA) and Givarlais (France).

A typical example is shown in Figure 8. For the best correlation that correspond to a temperature reversal of 483 s, the same as measured in Baton Rouge, the correlation coefficient was 0.63 from 100 to 250°C and 0.62 from 180 to 250°C.
Most of the peaks recorded in the slave glow curve up and down temperature correspond to the peaks cited in reference [15]. These peaks are at 130, 145, 180, 205°C according to this publication and appear on Figure7.
With chip H, two more heating cycles were done to check if some entangled electrons were left after one temperature excursion. Test H2 still had an acceptable correlation coefficient of 0.30 from 100 to 250°C and 0.44 from 180 to 250°C. Test H3 had a very poor correlation coefficient of 0.23 from 100 to 250°C and 0.09 from 180 to 250°C.

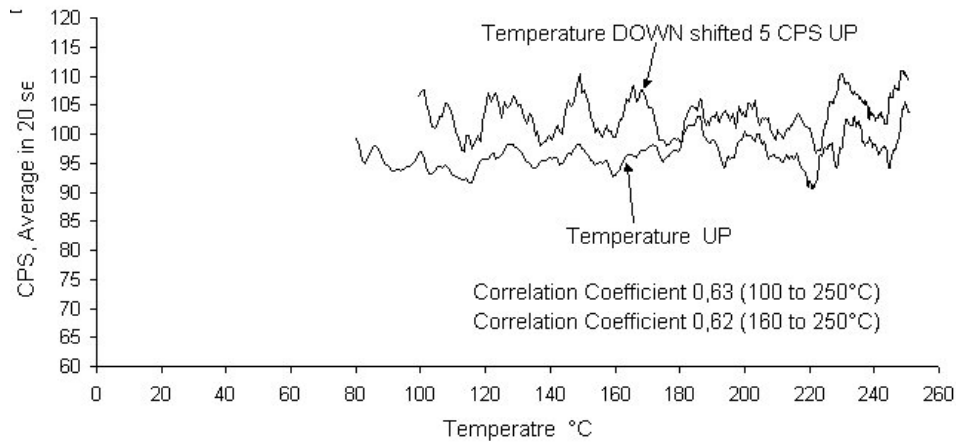

Figure 8 - Test of chips H1, LiF100, 20 Gray, PMT 2, September 16, 2006, Average in 20 s, Master sample heated in Baton Rouge, LA (USA), 8182 km, .Signal on Slave sample in Givarlais (France). Temperature reversal 483 s after start of oven heat.

Several tests were run with Chips LiF200. Figure 9 shows a typical run with chip D. The glow curves up and down correlate well both numerically and visually. Many peaks appear on these curves that have not been mentioned in the literature.

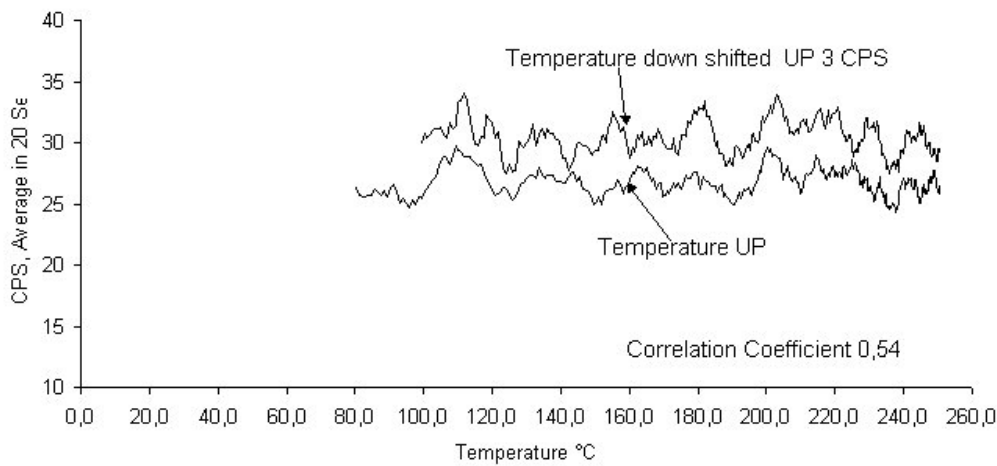

Figure 9 - Test of Chips D, LiF200, 10 Gray, PMT 1, September 2, 2006, average CPS in 20 s. Master sample heated in Baton Rouge, LA (USA), 8182 km. Signal on Slave sample in Givarlais (France). Temperature reversal 430 s after start of heat.

**9 - Conclusion**

The reported experiments are a practical implementation of the entanglement phenomenon of Quantum Mechanics. Two particles are said to be entangled when they are emitted simultaneously by the same atomic wave function, for example; photons emitted by a nucleus, or an electron, and the photons temporarily form an interactive interference pattern with one another. Such particles are quantum-*connected* to each other and interaction with a measurement system by one of them is "*sensed*" immediately by the entangled counterpart. Entanglement can be swapped between two particles and two other particles. Entangled particles, such as electrons, can be "stored" in ion traps or impurities within thermoluminescent crystal lattices and remain isolated from environmental decoherence effects in the traps for considerable amounts of time. Electrons can be forced to leave these traps and then drop down to their respective ground state energies in the crystal lattice by thermal heating or by stimulated luminescence. An entangled electron dropping out of its ion trap will go through spin transitions which affect its entangled counterpart electron by reason of spin conservation laws such that it becomes favorable for the counterpart electron to exit its trap as a result, emitting some light while dropping to ground state, at whatever distances the traps are located from one another. Since traps can be entangled even though present in separate crystal lattices, such samples can be separated by a large distance and the entangled electrons still be *connected* until perturbed by thermal heating of the crystal lattice containing one of the trapped entangled electron pairs. It appears that the trapped entangled electrons escape only at discrete and unique temperature values, thus allowing the same glow curve response (although much less intense than the heated crystal) to be recorded for each non-heated thermoluminescent crystal when the temperature of the heated crystal lattice is increased and decreased. This experiment amply demonstrates that:

- Bremsstrahlung gamma ray or X photons are entangled,
- The entangled photons can transfer their entanglement to particles (electrons)
- swapping entanglement between particles is possible and does occur,
- entangled particles can be "stored" as wave functions in ion traps that behave as QED cavities within thermoluminescent materials,
- environmental decoherence appears extremely feeble within ion traps containing entangled electrons since the heating and measurement experiments were conducted over one month after co-irradiation of separate TLD chips,
- entangled electrons appear to exit the traps only at very discrete and characteristic temperatures during temperature increase and not in accord with the Arrhenius equation which dictates that ordinary electron traps empty as a function of a continuum of release temperatures. This is a significant finding of this experiment which should provide quantitative clues for the interaction mechanisms involved in entangled electrons within ion traps
- slave chip (non-heated entangled counterpart crystal) glow curves correlate for the crystal lattice temperature increasing and then decreasing via cooling (temperature turn around point) in a very symmetrical and systematic way,
- quantum liaisons can be established between locations situated 8,182 km apart.

Due to the significance of these results, we would like to encourage other serious investigators to repeat these experiments. Please contact either or both of the authors via email or otherwise if you would like to replicate these experiments in order to obtain CLINAC irradiated TLDs, etc.

**Acknowledgment** :  The authors  thank E-Quantic Communications SARL-ACV for funding this research. Many thanks to the Centre de Radiothérapie Joseph Bellot of the Saint François clinic of Montluçon for irradiating the chips.